\documentclass[twocolumn,preprintnumbers,superscriptaddress,nofootinbib,aps,prd,floatfix,hyphens]{revtex4-2}
\pdfoutput=1

\usepackage{amsmath,amssymb}
\usepackage{wasysym}
\usepackage[pdftex]{graphicx}
\usepackage{epstopdf}
\usepackage{color,array,subfigure,slashed}
\usepackage[dvipsnames]{xcolor}
\usepackage{hyperref}

\newcommand{\be}{\begin{eqnarray}}
\newcommand{\ee}{\end{eqnarray}}

\newcommand{\bee}{\begin{eqnarray}}
\newcommand{\eee}{\end{eqnarray}}
\newcommand{\beeq}{\begin{equation}}
\newcommand{\eeeq}{\end{equation}}

\newcommand{\ifb}{{\text{fb}}^{-1}}

\numberwithin{equation}{section}

\begin{document}

\title{LHC constraints on electroweakino dark matter revisited}

\begin{abstract}
We revisit LHC searches for heavy invisible particles by exploiting QCD initial state radiation. We recast a dijet signal region in a general multijet plus MET search by ATLAS. We find that the nontrivial mass limit can be obtained for various models of the electroweakino sector with the present data in hadronic channels. The winos are bound to be heavier than $m_{\tilde W} \gtrsim  {160}$~GeV and Higgsinos $m_{\tilde h} \gtrsim  {100}$~GeV, depending on the chargino-neutralino mass splitting. The expected exclusion limits at the LHC Run 3 with $\mathcal{L} = 300\ \ifb$ increase to $m_{\tilde W} \gtrsim  {200}$~GeV and $m_{\tilde h} \gtrsim  {130}$~GeV for winos and Higgsino, respectively. This is the first LHC limit for promptly decaying nearly mass-degenerate winos.   

\end{abstract}

\author{Trygve Buanes} \email{Trygve.Buanes@hvl.no}
\affiliation{Western Norway University of Applied Sciences, Bergen, Norway\\[0.1cm]}

\author{I\~naki Lara} \email{Inaki.Lara@fuw.edu.pl}
\affiliation{Institute of Theoretical Physics, Faculty of Physics,\\University of Warsaw, ul.~Pasteura 5, PL--02--093 Warsaw, Poland\\[0.1cm]}

\author{Krzysztof Rolbiecki} \email{Krzysztof.Rolbiecki@fuw.edu.pl}
\affiliation{Institute of Theoretical Physics, Faculty of Physics,\\University of Warsaw, ul.~Pasteura 5, PL--02--093 Warsaw, Poland\\[0.1cm]}

\author{Kazuki Sakurai} \email{Kazuki.Sakurai@fuw.edu.pl}
\affiliation{Institute of Theoretical Physics, Faculty of Physics,\\University of Warsaw, ul.~Pasteura 5, PL--02--093 Warsaw, Poland\\[0.1cm]}

\preprint{}

\maketitle


\section{Introduction}
\label{sec:intro}

A variety of independent astrophysical and cosmological observations~\cite{Arbey:2021gdg} have grounded the inclusion in cosmological models of an electromagnetically neutral, nonbarionic matter species generically referred to as dark matter (DM). Despite a rich experimental effort to unveil the nature of dark matter via direct detection, collider searches, and indirect detection signals, it remains elusive.

In the case where DM has a substantial interaction with the Standard Model (SM), it can be produced at colliders. However, being electrically and color neutral, it will escape the detector without leaving any signal. Thus, an appropriate strategy to look for its production at collider experiments would be to measure an imbalance in a sum of transverse momentum of the visible particles traveling through a detector, when DM is produced in association with visible particles.

The signal where DM production is accompanied by at least one jet from initial state radiation (ISR)  has been considered conventionally as the most promising channel to detect DM at colliders. A number of dedicated searches have been performed, for recent ones see \cite{ATLAS:2021kxv,CMS:2021far}, which typically require 1--4 jets, with a very energetic leading jet,  and large missing transverse energy (MET). Early searches required exactly one~\cite{ATLAS-CONF-2011-096} or up to two jets~\cite{CMS-PAS-EXO-11-059} but it turned out that vetoing additional jets reduced acceptance~\cite{Dreiner:2012gx}. Nevertheless this strategy is usually called ``monojet".

Supersymmetry (SUSY) and in particular the Minimal Supersymmetric Standard Model (MSSM)~\cite{Haber:1984rc} has been a benchmark model for direct New Physics searches at the Large Hadron Collider (LHC). The lightest supersymmetric particle (LSP) is a DM candidate, though providing the right relic abundance is tricky~\cite{Bagnaschi:2015eha, Bagnaschi:2017tru,GAMBIT:2017zdo}. Here we focus on several specific scenarios where the LSP is the lightest neutralino, and is accompanied by charginos and heavier neutralinos, the supersymmetric partners of gauge and Higgs bosons~\cite{Haber:1984rc}.

The monojet strategy of Ref.~\cite{ATLAS:2021kxv} does not have enough sensitivity to constrain electroweakinos~\cite{Barducci:2015ffa} even around the LEP limit of 95--100~GeV~\cite{LEPlimits,Abdallah:2003xe,Abbiendi:2002vz,Acciarri:2000wy,Heister:2002mn}.\footnote{{Results of Ref.~\cite{Schwaller:2013baa} suggest it might be sensitive when 300 fb$^{-1}$ of data is collected.}} Therefore in this work we test sensitivity of another search strategy which was originally applied for squark and gluino pair production. As it turns out the main difference will be in the requirement of two energetic leading jets rather than one as for monojet.

\section{Scenarios\label{sec:scenarios}}

The realization of DM as a new species of weakly interacting massive particle (WIMP) has been hypothesized, justified by the feature of being able to reproduce the correct relic density for nonrelativistic matter in the early universe. A popular family of complete models including WIMPs is supersymmetry. Binos, Higgsinos and winos represent fermionic SU(2) singlets, doublets and triplets respectively, and the stability of the LSP is a consequence of the symmetries required to stabilize the proton. 

\begin{figure}[ht!]
 	\includegraphics[width=0.18\textwidth]{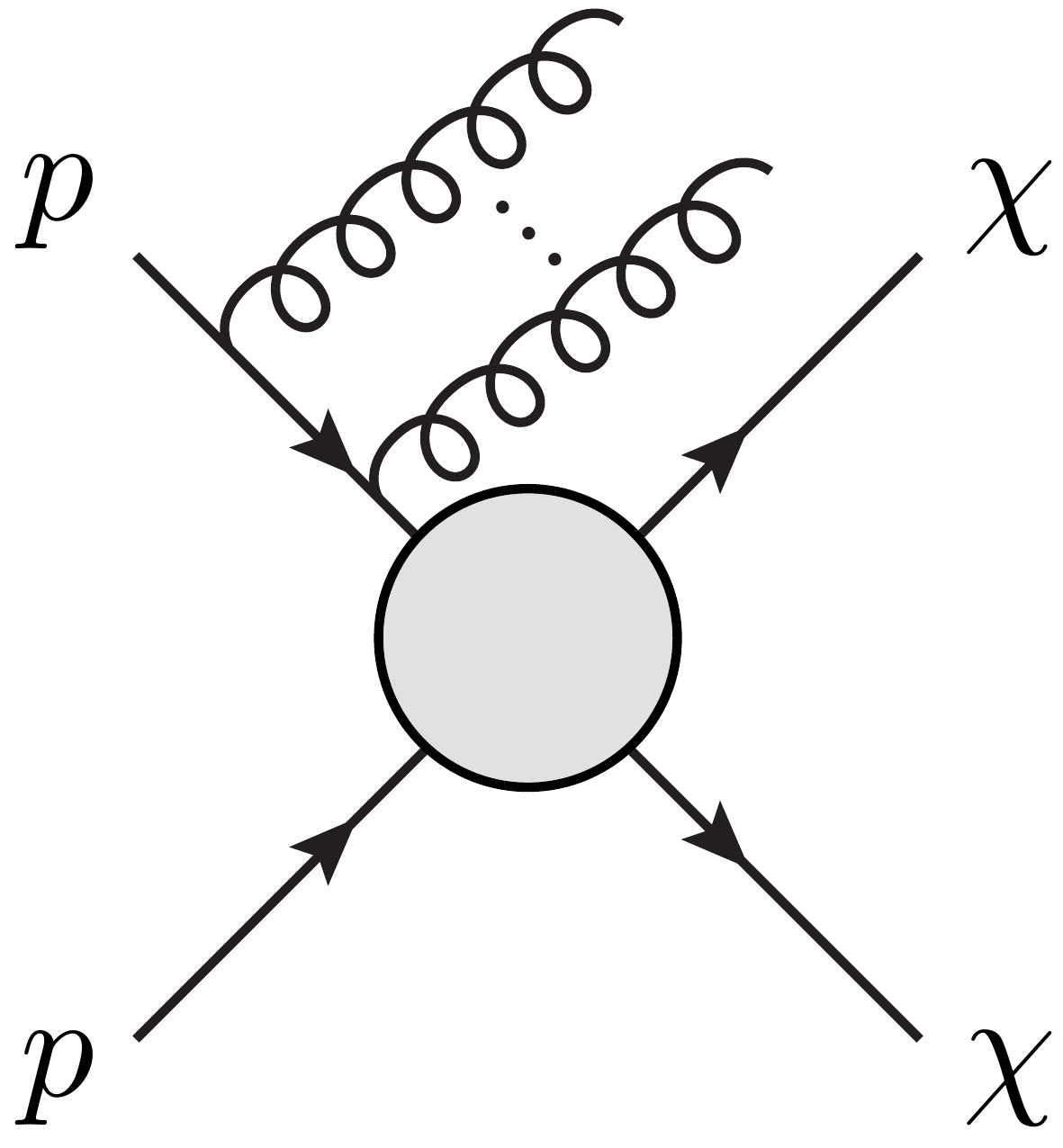}
 	\caption{Jets+MET signal from DM pair production at a hadron collider.}
 	\label{fig:DM_gen}
 \end{figure}

DM candidates charged under SU(2) can be directly produced at hadron colliders through $W$ and $Z$ mediated processes:
\begin{eqnarray}
&& u + \bar{d} \to \tilde{\chi}^+ + \tilde{\chi}^0 + X  \\
&& q + \bar{q} \to \tilde{\chi}^+ + \tilde{\chi}^- + X \\
&& q + \bar{q} \to \tilde{\chi}^0 + \tilde{\chi}^0 + X
\end{eqnarray}
where $X = 0j, 1j, 2j, ...$ denotes any number of additional jets, generically depicted in Fig.~\ref{fig:DM_gen}. In that case, the signal of a pair of DM particles with  an additional hard ISR jet can be strong enough to produce a distinguishable signal in jets+MET searches. 

\begin{figure*}[ht!]
    \includegraphics[width=0.3\textwidth]{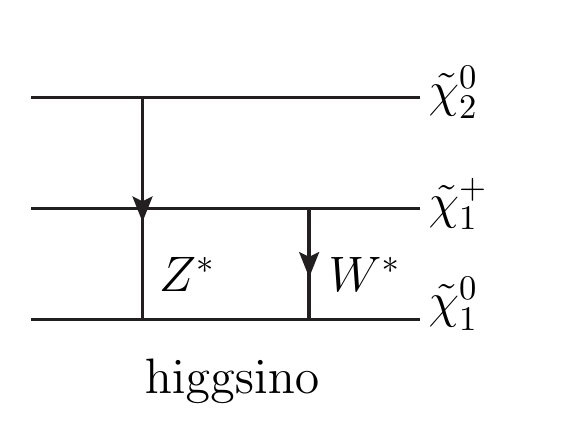}
 	\includegraphics[width=0.3\textwidth]{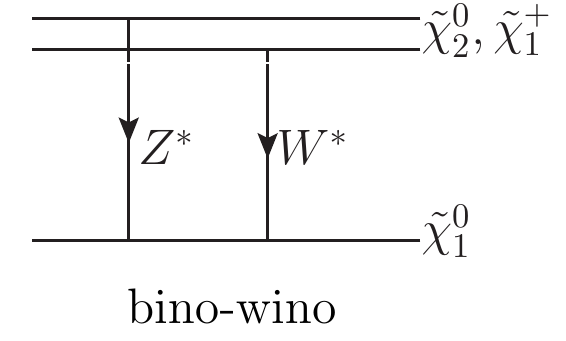}
 	\includegraphics[width=0.3\textwidth]{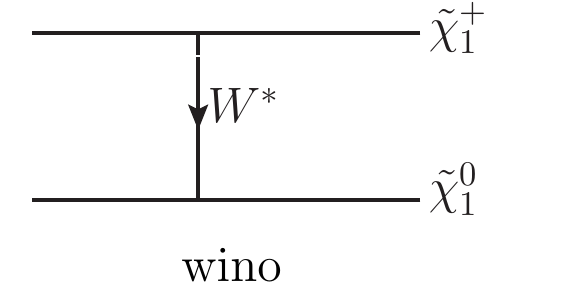}
 	\caption{The schematic spectra of the models discussed in Sec.~\ref{sec:scenarios} with main decay modes proceeding via off-shell electroweak gauge bosons.}
 	\label{fig:spectra}
 \end{figure*}

In the following, we are going to analyze the detection possibilities of WIMPs at the LHC, using the MSSM as a benchmark model with all of the spectrum decoupled, except for Higgsinos, winos or binos. In this simplified model, the collider phenomenology can be characterized by a small number of parameters (Higgsino and gaugino masses and $\tan \beta$). We are selecting three cases for study:
\begin{itemize}
    \item Higgsino LSP, heavy winos and bino:\\
    $\{ \tilde{\chi}^0_1, \tilde{\chi}^\pm_1, \tilde{\chi}^0_2\} \equiv \{\tilde{H}^0_1, \tilde{H}^\pm,  \tilde{H}^0_2\}$;
    \item bino LSP, light wino and heavy Higgsino:\\
    $\{ \tilde{\chi}^0_1, \tilde{\chi}^\pm_1, \tilde{\chi}^0_2\} \equiv \{\tilde{B}, \tilde{W}^\pm , \tilde{W}^0\}$;
    \item wino LSP, heavy Higgsinos and bino:\\
    $\{ \tilde{\chi}^0_1, \tilde{\chi}^\pm_1\} \equiv \{\tilde{W}^0, \tilde{W}^\pm \}$.
\end{itemize}
We refer to these scenarios by Higgsino, bino-wino and wino models, respectively. A characteristic feature of these models are small mass gaps between particles, which are the consequence of the mixing structure in the gaugino and Higgsino sectors, see e.g.~\cite{Martin:1997ns} for details. The full calculation within the MSSM must necessarily include higher order corrections~\cite{Pierce:1993gj, Ibe:2012sx}. In our simplified model setup, however, we will assume the mass gap to be a free parameter.  In the following we only consider mass gaps of the order 1~GeV but larger than 300~MeV. Generally the small mass differences will result in signatures specific to long-lived particles which are beyond the scope of this study. Figure~\ref{fig:spectra} shows schematically the particle spectra of the scenarios. 

In the Higgsino LSP model there are three states: $\tilde{\chi}^0_1$, $\tilde{\chi}^\pm_1$, $\tilde{\chi}^0_2$ with the following production modes:
\begin{equation}
    pp \to \tilde{\chi}^0_1 \tilde{\chi}^0_2,\ \tilde{\chi}^0_1 \tilde{\chi}^\pm_1,\ \tilde{\chi}^0_2 \tilde{\chi}^\pm_1,\ \tilde{\chi}^+_1 \tilde{\chi}^-_1. 
\end{equation}
The total production cross section of mass degenerate Higgsinos at $m_{\tilde H} = 100$~GeV is 18~pb at the NLO-NLL accuracy~\cite{Fuks:2012qx,  Fuks:2013vua}. For the mass gap of up to several GeV the LEP limits are between 92--102~GeV~\cite{Abbiendi:2002vz,Acciarri:2000wy,Heister:2002mn}.
At the LHC the ATLAS experiment has constrained Higgsinos above the LEP limit for the mass gap between neutralinos larger than 2 GeV~\cite{ATLAS:2019lng}.  Similarly CMS constraints the scenario with mass difference between neutralinos larger than 3~GeV~\cite{CMS:2021edw}.\footnote{The values lower than 3~GeV are not presented in the exclusion plot.} The constraints rely on the soft leptons arising in the decays of charginos and neutralinos via off-shell EW gauge bosons:
\begin{eqnarray}
&& \tilde{\chi}^\pm_1 \to W^* \tilde{\chi}^0_1 \label{eq:decayW},\\
&& \tilde{\chi}^0_2 \to Z^* \tilde{\chi}^0_1. 
\end{eqnarray}
An ISR jet provides a necessary boost to the chargino-neutralino system, so that the soft leptons can be detected above the SM background.\footnote{It was proposed that the Higgsino scenario could also be probed by a search for soft tracks~\cite{Fukuda:2019kbp} as well as in the mono-$Z$/$W$ channel~\cite{Carpenter:2021jbd}.} For the purpose of comparison with the experiments we assume the chargino mass to be halfway between neutralinos, i.e.\ $m_{\tilde{\chi}^\pm_1} = (m_{\tilde{\chi}^0_1} + m_{\tilde{\chi}^0_2})/2$ though in realistic full-model calculations the chargino mass tends to be closer to the heavier neutralino~\cite{Porod:2003um,Porod:2011nf}. In any case, this results in a kinematic suppression of a decay mode $\tilde{\chi}_2^0 \to \tilde{\chi}^\pm_1 + X$.

In the bino-wino model there are three states ordered with a growing mass: $\tilde{\chi}^0_1$, $\tilde{\chi}^\pm_1$, $\tilde{\chi}^0_2$. The winos are mass degenerate, while the mass gap between the bino and the winos, $\Delta m$, is of the order of several GeV.\footnote{The second lightest neutralino, $\tilde \chi_2^0$, tends to be long-lived in this scenario depending on the Higgsino mass \cite{Rolbiecki:2015gsa,Nagata:2015pra}.  We do not consider such a case.}
Owing to the coannihilation mechanism, such a setup can in principle provide a correct relic DM abundance~\cite{Baer:2005jq,Arkani-Hamed:2006wnf,Ibe:2013pua,
Harigaya:2014dwa}. The following production modes are available:
\begin{equation} \label{eq:winoprod}
    pp \to \tilde{\chi}^0_2 \tilde{\chi}^\pm_1,\ \tilde{\chi}^+_1 \tilde{\chi}^-_1. 
\end{equation}
The total production cross section of mass degenerate winos at $m_{\tilde W} = 100$~GeV is 34.3~pb at the NLO-NLL accuracy~\cite{Fuks:2012qx,  Fuks:2013vua}.
The production modes involving bino vanish. The general limits from LEP at around 100 GeV apply here as well. At the LHC, this scenario is excluded for a mass difference of 1.8 GeV between bino and winos at the mass of 100 GeV~\cite{ATLAS:2019lng}. For the mass difference of 10~GeV, the limit goes as far as 270~GeV~\cite{CMS:2021edw}.

Finally for the wino model there are two states, $\tilde{W}^0$ and $\tilde{W}^\pm$, with a mass difference $\Delta m = \mathcal{O}(\mathrm{GeV})$. The production modes are the same as in the bino-wino model. The LEP constraints are in a range 92--102 GeV for $\Delta m < 10$~GeV. No LHC constraints existed to date.

\section{Methods}

For each point of the parameter corresponding to the supersymmetric scenarios, with the spectrum calculated as explained in the previous section, we proceed as follows: We generate Monte Carlo events using \texttt{MadGraph5\_aMC@NLO 3.1.0}~\cite{Alwall:2014hca,Alwall:2007fs,Alwall:2008qv} for the hard process corresponding to the inclusive production of a pair of Higgsinos 
or winos 
plus the emission of up to two partons at matrix element level. A generator-level cut is imposed requiring at least one jet with 200 GeV of transverse momentum and the  NNPDF23LO~\cite{Ball:2012cx,Buckley:2014ana} PDF set is used. The events are then interfaced to \texttt{Pythia 8.244}~\cite{Sjostrand:2014zea} for modeling of the Higgsino/wino decay, parton showering and hadronization. Jet clustering is performed using \texttt{fastjet 3.3.4}~\cite{Cacciari:2011ma,Cacciari:2005hq,Cacciari:2008gp}. Jet matching and merging to parton-shower calculations is accomplished by the MLM algorithm~\cite{Mangano:2006rw}. The inclusive cross section obtained is rescaled using the $K$-factor corresponding to the exclusive hard process without ISR emission calculated using \texttt{Resummino 2.0.1}~\cite{{Debove:2011xj,Debove:2009ia,Debove:2010kf,Fuks:2013vua,Fuks:2012qx,Fiaschi:2018xdm,Fiaschi:2020udf}} at NLO+NLL.
The resulting events are fed to \texttt{CheckMATE}~\cite{Dercks:2016npn,Kim:2015wza,cm2} to test against the present limits from ATLAS monojet~\cite{ATLAS:2021kxv} and multijet~\cite{ATLAS:2020syg} searches. \texttt{CheckMATE} is a universal tool for recasting of LHC searches in context of arbitrary New Physics models. It uses the fast detector simulation framework \texttt{Delphes}~\cite{deFavereau:2013fsa} with customized ATLAS detector card and additional built-in tuning for a more accurate reproduction of experimental efficiencies.

In addition, we  implemented a shape fit over the multi-bin signal regions \texttt{MB-SSd} and \texttt{MB-C} defined in the multijet analysis~\cite{ATLAS:2020syg}, using the package \texttt{pyhf}~\cite{Cowan:2010js,pyhf_joss,pyhf}, which is a Python implementation of the \texttt{HistFactory} specification for binned statistical models~\cite{Cranmer:1456844,Baak:2014wma}, and following recommendations of Ref.~\cite{ATL-PHYS-PUB-2021-038}. This method turns out to be superior to the usual limits calculated by \texttt{CheckMATE} using only the expected most sensitive signal region~\cite{Read:2002hq}. 

The bins in the multibin signal regions are by construction orthogonal and therefore allow for the statistical combination. This can be performed in a simplified way when the background model is approximated by the total SM background rate set to the post-fit background rate and uncertainty obtained in the background-only full likelihood fit provided in the auxiliary material. The signal strength $\mu$ is introduced as an unconstrained parameter in the likelihood and it is the parameter of interest used for hypothesis tests and model exclusion~\cite{ATL-PHYS-PUB-2021-038}. This approximation assumes that background uncertainties are controlled by a single shape-uncertainty parameter correlated over all bins. This simplified method is used for efficient scanning of the parameter space. Additionally, we validate these results using the full likelihood model provided by ATLAS. The full likelihood method is significantly more complex with many additional bin-wise numbers representing different background contributions, many additional nuisance parameters and accounting for the full background correlations. We observe a good agreement between both approaches. A more detailed comparison of these two methods for a number of searches will be provided in forthcoming publication; see also Ref.~\cite{Alguero:2022gwm} for more discussion. 

The method of kinematical shape-fit described in the preceding paragraphs offers a significant advantage over the single most sensitive signal region method. Further examples can be found in Refs.~\cite{ATL-PHYS-PUB-2021-038} and \cite{Alguero:2022gwm}. Nothing of the parameter space of our model of interest would be excluded in that way. For the signal regions \texttt{MB-SSd} and \texttt{MB-C} we also note that while they offer similar \textit{expected} excluding power, the observed limits are stronger in \texttt{MB-SSd} due to occasional positive background fluctuations in some of the bin of  \texttt{MB-C}.  

\begin{figure}[ht!]
\includegraphics[width=0.45\textwidth]{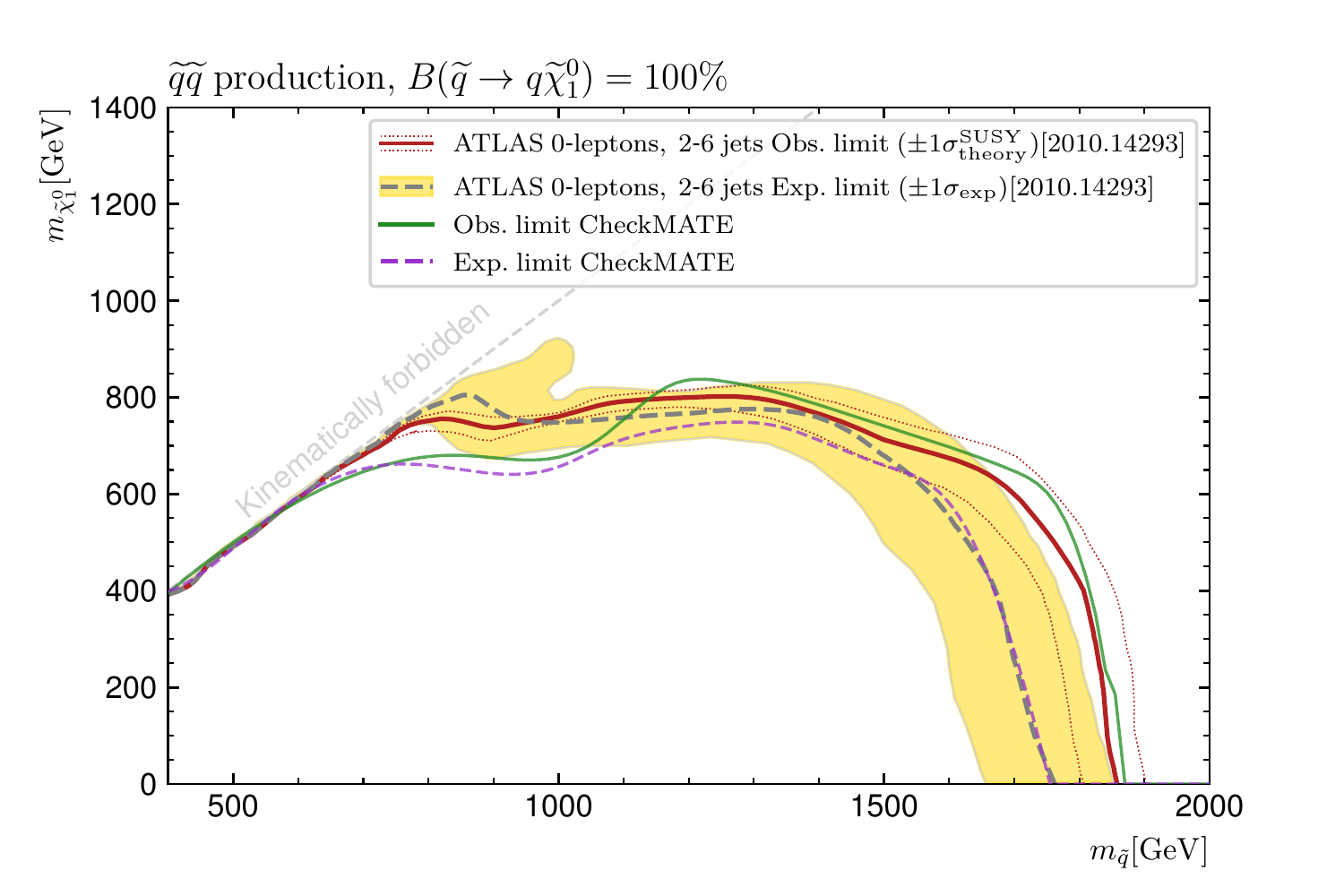}
\caption{The exclusion in the squark-LSP mass plane obtained by the ATLAS experiment~\cite{ATLAS:2020syg}: expected -- grey dashed line and observed -- red solid line; and obtained with \texttt{CheckMATE} using the simplified method: expected -- purple dashed line and observed -- green solid line. Note that for the sake of a direct test of the \texttt{MB-SSd} exclusion power, the  \texttt{CheckMATE} result uses just the \texttt{MB-SSd} signal region, while the ATLAS results also includes the \texttt{MB-GGd} and \texttt{MB-C} signal regions. This results in the weaker than expected exclusion in the compressed-mass region. \label{fig:validation_SSd}  }
\end{figure}

In order to document validation of our implementation we include the exclusion contour for the squark pair production using only the \texttt{MB-SSd} channel which is relevant for this study, Fig.~\ref{fig:validation_SSd}. The plot compares the expected and observed exclusion obtained by ATLAS, dashed grey and solid red curves, respectively, with expected and observed exclusion obtained with \texttt{CheckMATE} using the simplified method discussed above. A very good agreement is observed for $m_{\tilde{q}} \gtrsim 1100$~GeV, while for lighter squarks the \texttt{CheckMATE} exclusion is visibly weaker. This is due to the fact that the full ATLAS results include also the \texttt{MB-GGd} and \texttt{MB-C} channels which have better expected limits in the squark compressed-mass region.  Since in our case the strongest exclusion comes from the \texttt{MB-SSd} signal region we restrict the plot to show that there is no significant additional uncertainty due to the recasting procedure.   

Following this prescription, we test masses of the LSP in the range 60--160 GeV and values of the mass gap between the LSP and the NLSP in the range 0.4--20 GeV. Below 0.4~GeV the searches for long-lived massive particles become sensitive, especially long-lived charginos that travel a microscopic distance in the detector resulting in the ``disappearing track" signature~\cite{ATLAS:2022rme}. As expected, the monojet search of Ref.~\cite{ATLAS:2021kxv} is weaker than the multijet search over the whole tested parameter space. In the multijet search the sensitivity comes from the shape fit in the \texttt{SSd} signal regions with 2 or 3 jets. The key feature of the \texttt{SSd} signal regions is the requirement that for the two leading jets $p_T(j)>250$~GeV and $\mathrm{MET} > 300$~GeV.

An uncertainty in the cross section calculation is estimated in {\tt MadGraph5} by varying factorization, renormalization and emission scales in the range $[0.5-1.5]$ around the central values. The largest contribution comes from the emission scale variation: $^{+20\%}_{-25\%}$. PDF uncertainty of $2.4\%$ was subdominant and the cross section was stable within a few \% against the choice of different dynamic scales schemes as implemented in {\tt MadGraph5}.

\section{Results}
\label{sec:results}

\begin{figure*}[h!tbp]
	\includegraphics[width=0.48\textwidth]{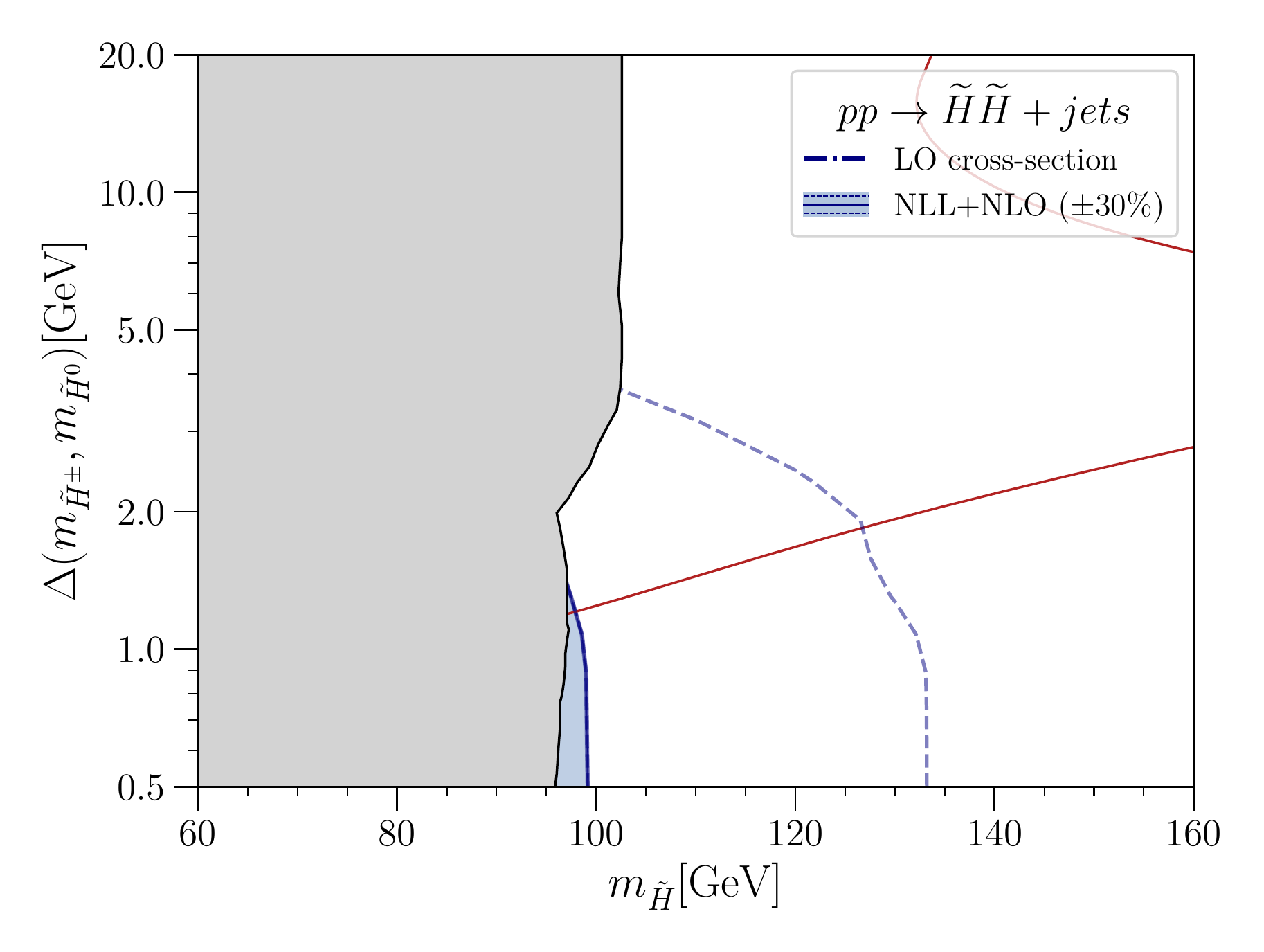}
	\includegraphics[width=0.48\textwidth]{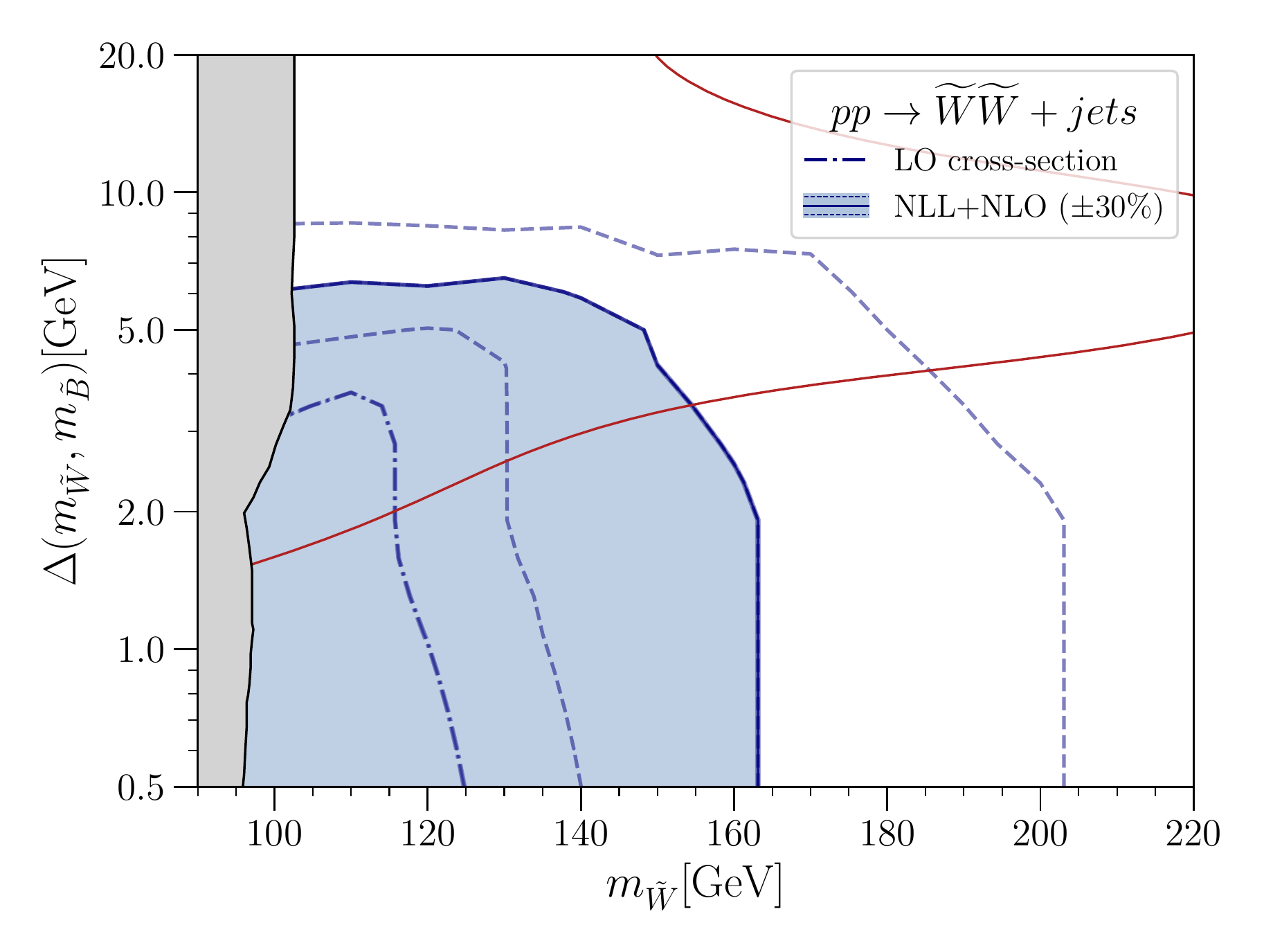}\\
	\includegraphics[width=0.48\textwidth]{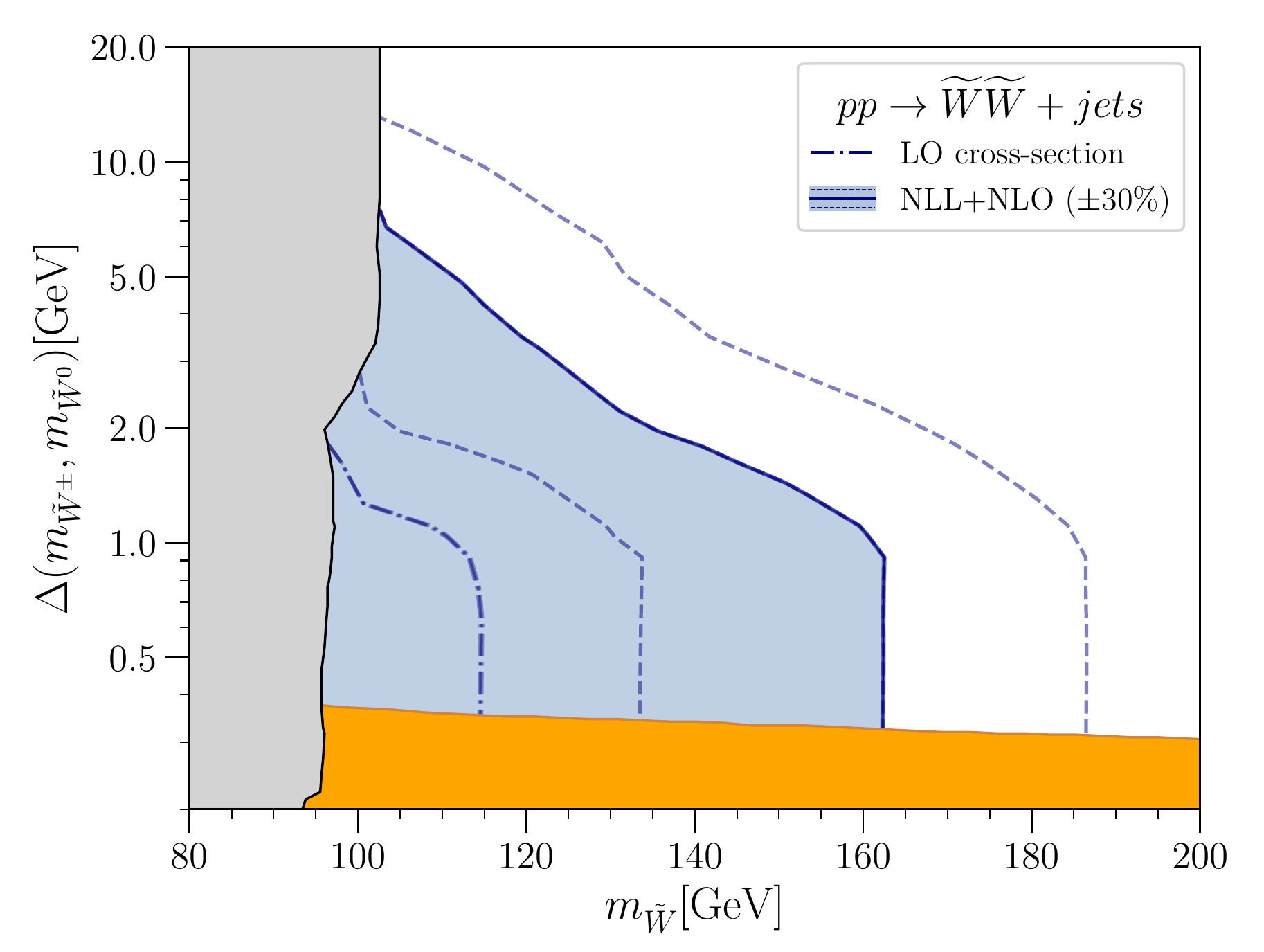}
	\includegraphics[width=0.48\textwidth]{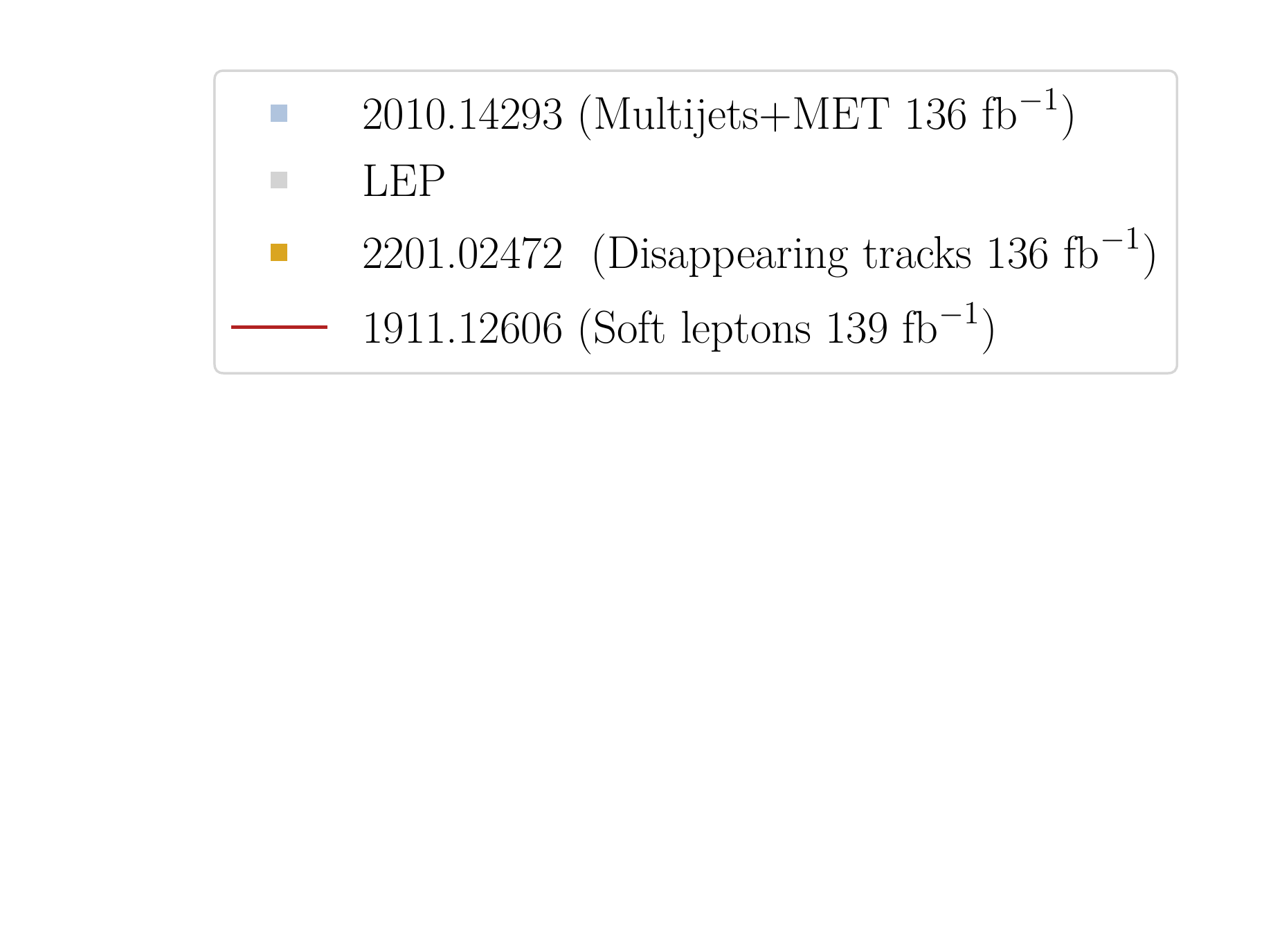}
	\caption{The upper left, upper right and lower left plots represent the limits on the [$m_{\widetilde{H}},~\Delta(m_{\tilde{H}^\pm},m_{\tilde{H}^0})]$ plane of the Higgsino model, [$m_{\tilde{W}},~\Delta(m_{\tilde{W}^\pm},m_{\tilde{B}^0})$] plane of the wino-bino model and [$m_{\tilde{W}},~\Delta(m_{\tilde{W}^\pm},m_{\tilde{W}^0})$] plane of the wino model, respectively. Blue area is excluded by ATLAS multijet search~\cite{ATLAS:2020syg} recasted in this work and the uncertainty from the cross section (NLO+NLL) calculation and recasting is denoted by the blue-dashed lines. An alternative exclusion assuming LO cross section is denoted with the blue dash-dotted line. The gray area is excluded by the combination of LEP data~\cite{Abbiendi:2002vz,Acciarri:2000wy,Heister:2002mn}. The area inside of the red line is excluded by ATLAS soft lepton searches~\cite{ATLAS:2019lng} and the orange area is excluded by the disappearing track search~\cite{ATLAS:2022rme}.
	}
	\label{fig:EW_res}
\end{figure*}

Following the strategy discussed in the previous section, we obtained exclusions for both the gaugino and Higgsino scenarios and these are shown in Fig.~\ref{fig:EW_res}. The horizontal axis corresponds to the LSP mass, while the vertical axis to the characteristic mass splitting for each model. In the plots we also include the best limits obtained using the soft lepton search by the ATLAS experiment~\cite{ATLAS:2019lng} (the red line). The LEP limits~\cite{LEPlimits,Abdallah:2003xe,Abbiendi:2002vz,Acciarri:2000wy,Heister:2002mn} are also shown as the grey shaded area and for the wino model we include the result of search for long-lived charginos~\cite{ATLAS:2022rme} (orange shaded). In all plots the exclusion obtained using the $K$-factor corrected cross section is denoted with a blue solid line. Blue dashed lines denote a $\pm 30\%$ error band on the central exclusion line and cover emission scale uncertainty, PDF uncertainty and recasting uncertainty. Apart from that, the blue dash-dotted line shows the exclusion when the cross section is calculated at the leading order by {\tt MadGraph5}.  

\begin{figure*}[h!tbp]
	\includegraphics[width=0.48\textwidth]{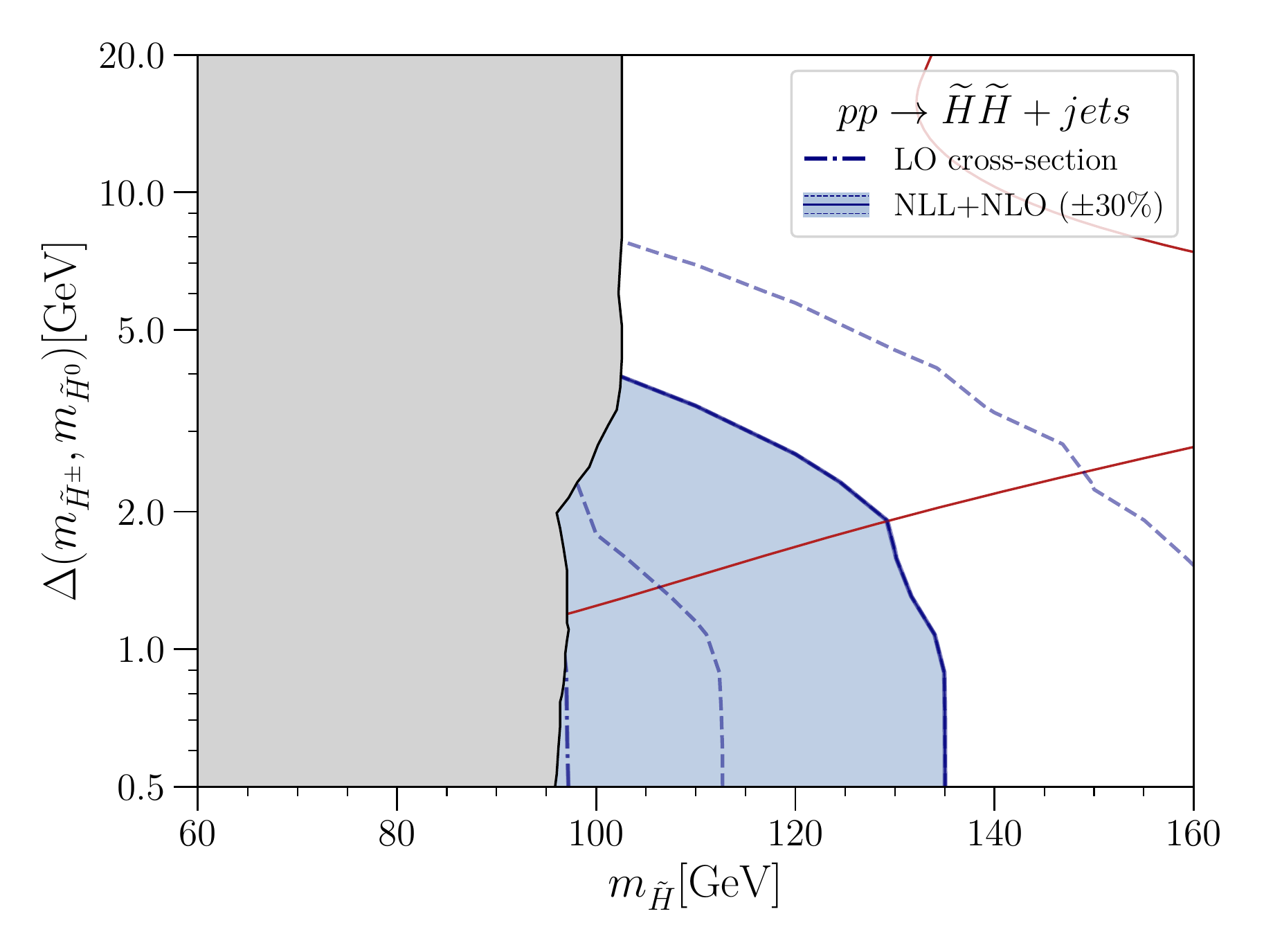}
	\includegraphics[width=0.48\textwidth]{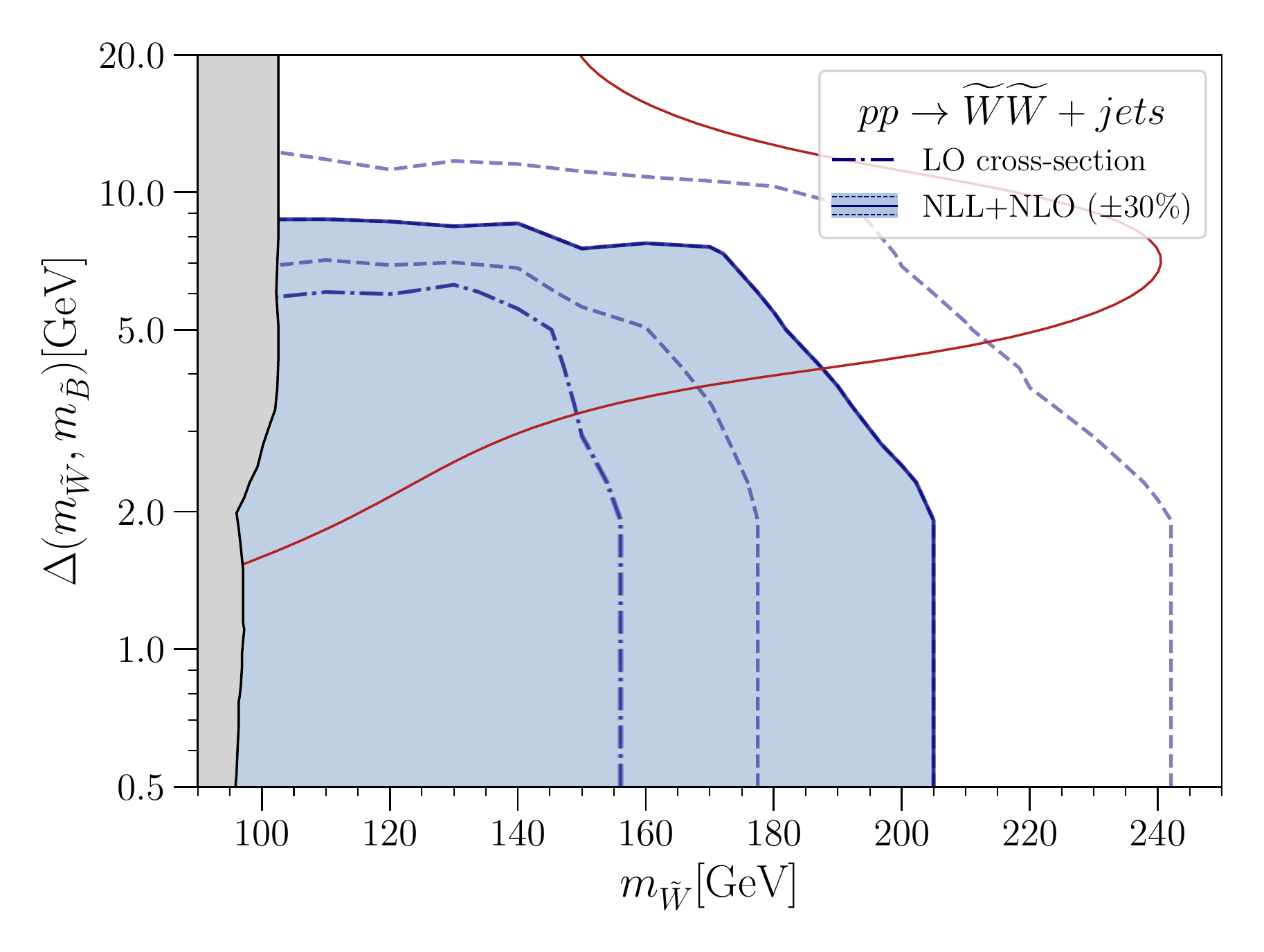}\\
	\includegraphics[width=0.48\textwidth]{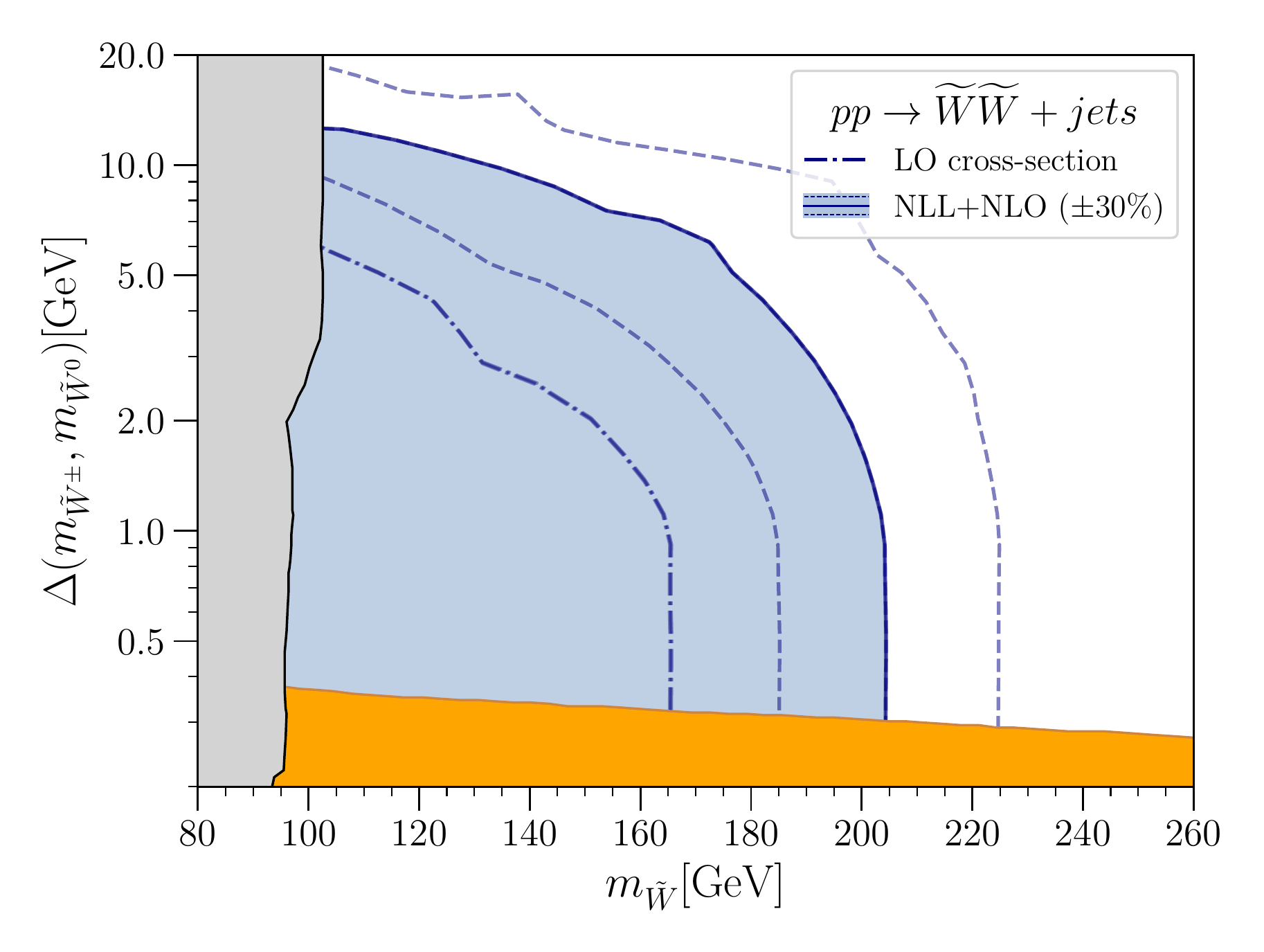}
	\includegraphics[width=0.48\textwidth]{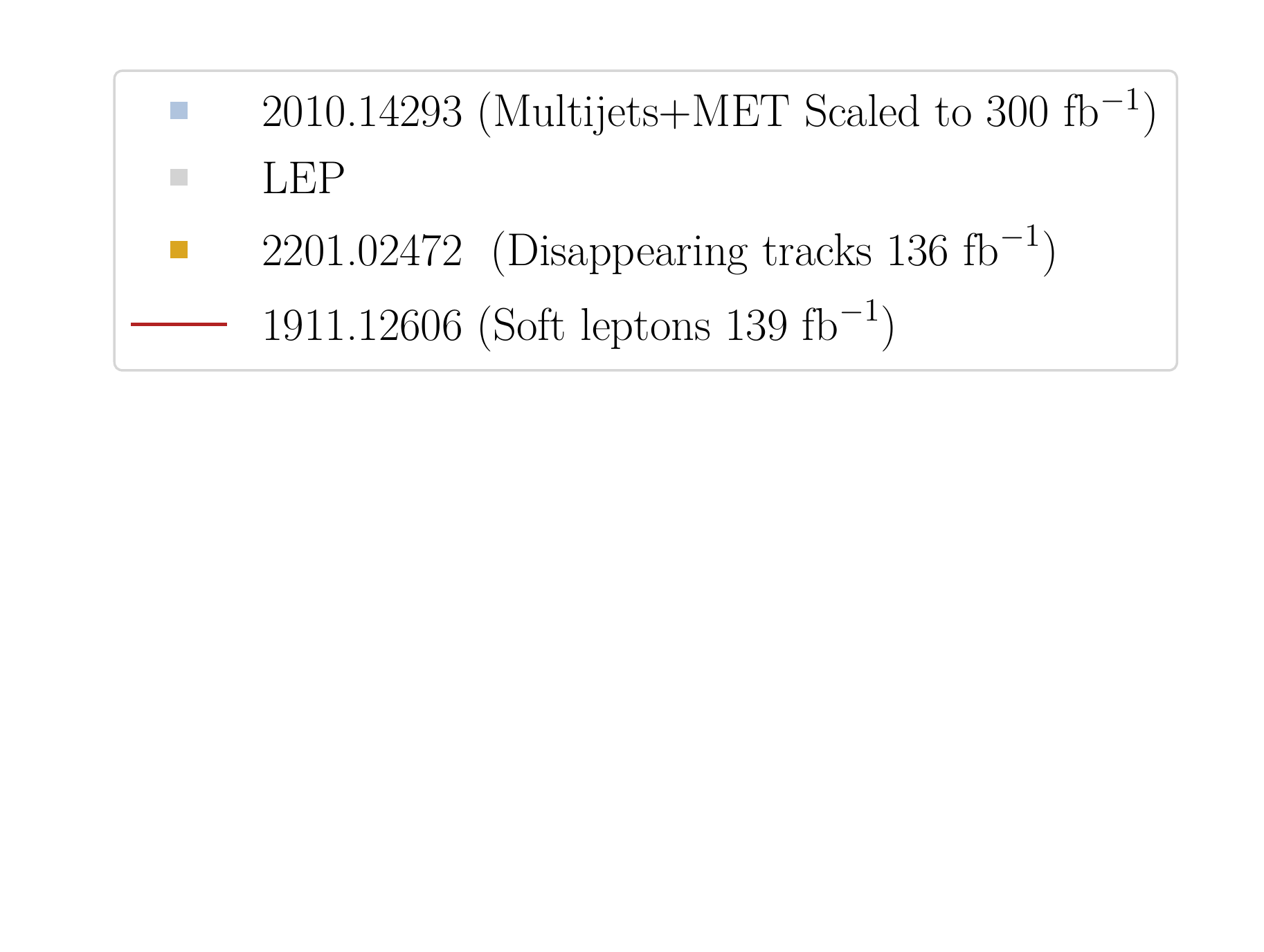}
	\caption{Same as Fig.~\ref{fig:EW_res} but for LHC Run 3 with 
	$\mathcal{L} = 300\ \ifb$.
	\label{fig:EW_res_HL}}
\end{figure*}

The weakest constraint is derived for the Higgsino model, see the upper left panel of Fig.~\ref{fig:EW_res}. At the mass splitting $\Delta m = 0.5$~GeV we exclude Higgsinos with the masses up to 100 GeV, which just slightly improves the LEP limit of 95 GeV. This model exhibits the lowest cross section which amounts to $\sigma \cdot K =220~\mathrm{fb} \cdot 1.5 = 330$~fb, where $\sigma$ is the cross section calculated by {\tt MadGraph5} for the pair production of higgsinos with up to two additional partons and $p_T > 200$~GeV for the leading parton, while $K$ is the scaling factor calculated using \texttt{Resummino} at the NLO-NLL order.  

For the bino-wino model, the upper right panel of Fig.~\ref{fig:EW_res}, we significantly extend previous limits. Wino with a mass up to 158 GeV and wino-bino mass difference 2 GeV are excluded. The total effective cross section for the winos of mass 100~GeV is much higher compared to the Higgsino scenario: $\sigma \cdot K = 425~\mathrm{fb} \cdot 1.55 = 660$~fb, obtained as explained above. For the mass difference $\Delta m > 6$~GeV the sensitivity is lost because of the selection requirement
\begin{equation}\label{eq:azimuth}
    \Delta \phi(j, \mathbf{p}_T^\mathrm{miss}) > 0.8,
\end{equation}
i.e.\ that the azimuthal distance between three leading signal jets and the missing transverse momentum is larger than 0.8 radians. This can be understood in the following way: the decay products of winos, albeit generally soft, will be highly boosted due to the presence of ISR. The direction of the boost will align with the boost of the LSPs system and hence with the direction of $\mathbf{p}_T^\mathrm{miss}$. With the increasing mass splitting the jet formed by the decay products will be increasingly more likely to pass a threshold of 50 GeV and become classified as one of the signal jets violating selection Eq.~\eqref{eq:azimuth}.  

Finally, in the lower left panel of Fig.~\ref{fig:EW_res} we present a result for the wino scenario. Winos up to the mass of 160 GeV are excluded for the mass difference $\Delta m < 1$ GeV. At $m_{\tilde W} \sim 110$ GeV, on the other hand, the exclusions goes up to $\Delta m \sim 7$~GeV. The only competing exclusion, apart from the LEP results,  comes from the long-lived chargino search exploiting the disappearing track signature~\cite{ATLAS:2022rme}. The soft lepton searches are not sensitive to this scenario as they require a pair of same-flavour opposite-charge leptons with a low invariant mass originating from the neutralino decays. Here, a pair of leptons can only be produced in chargino pair production, cf.\ Eq.~\eqref{eq:winoprod}, and the following decay mediated via on off-shell $W$, Eq.~\eqref{eq:decayW}. Hence the limit presented here is the only existing constraint at the LHC for promptly decaying winos.  Similarly to the bino-wino case, the sensitivity of the multijet search decreases for a larger mass difference within the wino system. 

Additionally, in Fig.~\ref{fig:EW_res_HL} we provide a projection of exclusion limits expected with additional luminosity collected during Run 3. For this purpose the expected signal and background numbers along with background uncertainty were scaled proportionally. Since with more data collected one may expect more accurate background predictions this projections should be regarded as conservative. We observe that the exclusion limits at their maximum extent increase to $m_{\tilde h} \gtrsim  {132}$~GeV, $m_{\tilde W} \gtrsim  {208}$~GeV and  $m_{\tilde W} \gtrsim  {204}$~GeV for the Higgsino, bino-wino and wino model, respectively.

\section{Conclusions}
\label{sec:conclusions}
We have proposed that the existing searches at the LHC in the multijet+MET final state can be applied to the challenging case of compressed MSSM electroweakino spectra near the electroweak scale. We recasted the ATLAS analysis and obtained new exclusion limits when the mass difference between the LSP and NLSP is below  5~GeV. For the Higgsino LSP the limit improves only slightly over the existing LEP limit, while  for the bino LSP accompanied by winos we exclude the respective wino up to a mass of 160~GeV for the mass differences up to 2~GeV. In the pure wino model, the winos are excluded up to 160 GeV for the mass difference of 1~GeV. This fills an important gap in the coverage of electroweakino spectra. 
We also provided the extrapolation to higher luminosity expected after Run 3, with the limits increasing by several tens of GeV. This information can serve to guide studies of physics prospects at future colliders.

\medskip

\noindent{\it{Acknowledgements}} --- 
The research leading to these results has received funding from the Norwegian Financial Mechanism 2014-2021, grant 2019/34/H/ST2/00707. The work of I.L.\ was directly funded from the grant. The work of K.S.\ is partially supported by the National Science Centre, Poland, under research grant 2017/26/E/ST2/00135. K.R.\ was partially supported by the National Science Centre, Poland under grants: 2018/31/B/ST2/02283, 2019/35/B/ST2/02008.


\bibliography{references.bib}

\end{document}